\documentclass[aps,prl,twocolumn,groupedaddress]{revtex4-1}
\usepackage{amsmath}
\usepackage{graphicx}
\usepackage{color}
\begin{document}
\title{Ultrafast pseudospin dynamics in graphene}
\author{M. Trushin$^1$, A. Grupp$^1$, G. Soavi$^{1,2}$, A. Budweg$^1$, D. De Fazio$^2$, U. Sassi$^2$,
A. Lombardo$^2$, A. C. Ferrari$^2$, W. Belzig$^1$, A. Leitenstorfer$^1$, and D. Brida$^1$}
\email{daniele.brida@uni-konstanz.de}
\affiliation{$^1$Department of Physics and Center for Applied Photonics, University of Konstanz, D-78457 Konstanz, Germany\\
$^2$Cambridge Graphene Centre, University of Cambridge, Cambridge CB3 0FA, UK}

\begin{abstract}
Interband optical transitions in graphene are subject to pseudospin selection rules. Impulsive excitation with linearly polarized light generates an anisotropic photocarrier occupation in momentum space that evolves at timescales shorter than 100fs. Here, we investigate the evolution of non-equilibrium charges towards an isotropic distribution by means of fluence-dependent ultrafast spectroscopy and develop an analytical model able to quantify the isotropization process. In contrast to conventional semiconductors, the isotropization is governed by optical phonon emission, rather than electron-electron scattering, which nevertheless contributes in shaping the anisotropic photocarrier occupation within the first few fs.
\end{abstract}
\maketitle
{\em Introduction---} The unique optical and electronic properties of graphene are appealing for advanced applications in photonics and optoelectronics\cite{Bonaccorso2010review,Nanoscale2015roadmap}. A variety of prototype devices have already been demonstrated, such as transparent electrodes in displays\cite{Natnano2010bae} and photovoltaic modules\cite{ACSnano2010dearco}, optical modulators\cite{Nature2011liu}, plasmonic devices\cite{Nature2011liu,Natnano2011ju,Natcomm2011echtermeyer,ACSnano2010schedin,Nature2012fei,Nature2012chen}, microcavities\cite{Natcomm2012engel,Nanolett2012furchi} and ultra-fast lasers\cite{ACSnano2010sun}. Amongst these, a significant effort is being devoted to the development of broadband photodetectors\cite{Natnano2014review}.
In this context, the same properties that make this two-dimensional (2d) material so appealing are also profoundly different from standard semiconductors. Thus, in order to fully exploit the technological potential of graphene, a full and comprehensive description of the physical phenomena occurring when light excites transitions within the Dirac cones is needed.

The ultrafast carrier dynamics has been extensively investigated in graphene\cite{PRL2010lui,PRB2011breusing,PRB2011hale,Natcomm2013brida,ACSnano2011shang} by tracking the evolution of the non-equilibrium distribution created by impulsive optical excitation. Once a Fermi-Dirac distribution is established\cite{Natcomm2013brida,PRB2013tomadin}, it cools down by optical phonon emission at the hundred fs timescale\cite{PRB2011breusing,PRL2005lazzeri} and by scattering on acoustic phonons\cite{PRL2012song,PRL2009bistritzer,PRB2009tse} at the ps timescale. However, the complete description of the phenomena responsible for the momentum-space photocarrier redistribution within the first 100fs requires significant efforts: a complex interplay between electron-electron (e-e)\cite{Natcomm2013brida,PRB2013tomadin} and electron-phonon (e-ph)\cite{APL2012malic,PRL2005lazzeri} scattering dominates the thermalization dynamics.

\begin{figure}
\includegraphics[width=\columnwidth]{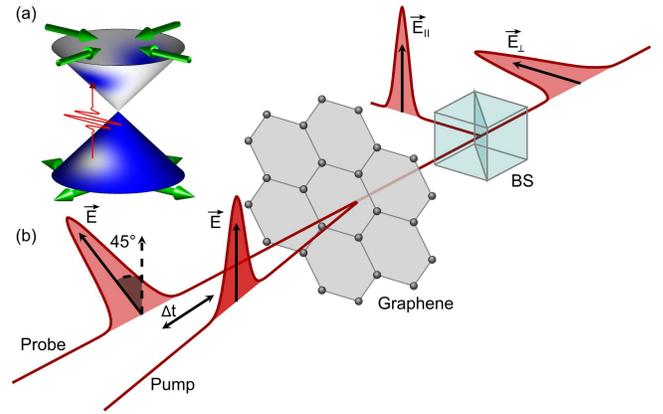}
\caption{\label{Fig1}
(a) Anisotropic carrier distribution after excitation with linearly polarized light due to pseudospin-momentum coupling. (b) A linearly polarized pump pulse is followed, after time delay $\Delta t$, by a probe pulse with polarization rotated by $45^\circ$. After interaction with graphene, a polarizing beam splitter separates parallel (E$_\parallel$) and perpendicular (E$_\perp$) components of the probe field for simultaneous detection.}
\end{figure}

Several theoretical\cite{PRB2011malic,APL2012malic} and experimental\cite{Nanolett2014mittendorff,
Nanolett2014echtermeyer} studies demonstrated the possibility to generate an anisotropic photocarrier distribution in the momentum space by means of linearly polarized light, and exploiting the pseudospin selection rules, which follow from the two sublattices present in graphene\cite{Nature2007geim}. The pseudospin can be seen as an internal angular momentum\cite{EPL2012trushin}. Thus, similarly to a gyroscope, a torque has to be applied in order to change the pseudospin orientation\cite{EPL2012trushin}. This can be achieved by the interaction with an electromagnetic wave whose electric field provides the necessary momentum\cite{EPL2012trushin}. The torque equals zero when the pseudospin is parallel to the electric field polarization making its flip forbidden. In contrast, this process is most efficient when the light polarization and pseudospin are normal to each other\cite{EPL2012trushin}, and the torque is maximized. Thus, as shown in Fig.\ref{Fig1}(a), a pseudospin flip is necessary to directly excite an 
electron from the valence to the conduction band. Hence, an excitation with linearly polarized light results in an anisotropic carrier occupation within the Dirac cone.

Here we measure the real-time isotropization of the photocarrier distribution in graphene for different concentrations of the out-of-equilibrium carriers and explain the main mechanism responsible for the anisotropy relaxation. We perform fluence-dependent measurements of the carrier anisotropy with few-fs temporal resolution and develop an analytical description of the transient optical absorption. This allows us to resolve the pseudospin dynamics over its timescale. The combination of experiment and theory allows us to demonstrate that (i) the photocarrier distribution anisotropy is stronger for lower radiation intensities when the anisotropic character of the excitation is not yet suppressed by the isotropic Fermi-Dirac sea; (ii) the isotropization dynamics is driven by the carriers' scattering with optical phonons, whereas e-e scattering is mainly responsible for the initial photocarrier redistribution along the Dirac cone. The latter observation is in stark
contrast with the case of conventional direct-gap semiconductors, like GaAs\cite{JETP1991merkulov},
where the anisotropies are typically lost via carrier-carrier scattering\cite{JoAP2004schneider,PRL2014kanasaki}.

{\em Experiment---} We study a chemical vapor deposited single layer graphene (SLG)\cite{Materials2014bonaccorso,Natnano2010bae} transferred onto a thin fused silica substrate, as described in Appendix. Structural quality, uniformity and doping of SLG before and after transfer are investigated by Raman spectroscopy\cite{PRL2006ferrari,Natnano2013ferrari}. This shows that no damage occurs to the sample as a consequence of the transfer process. The Raman measurements indicate that the sample is p doped, 
with a Fermi level$\sim$200meV\cite{NatNano2008das,PRB2009basko}. The D to G ratio measured at 514.5nm is $\sim$0.24, which, combined with the estimated doping, corresponds to a small defect concentration$\sim$10$^{11}$cm$^{-2}$\cite{ACSNano2014bruna}.

Transient absorption measurements are done with a 15fs pump pulse at central photon energy of 1.62eV ($\lambda$=765nm), see Appendix. A degenerate configuration is employed, where the same pulse is split and used for both excitation and probing. The experiments are performed as shown in Fig.\ref{Fig1}b. The polarization of the probe pulse is rotated by 45$^{^o}$ with respect to the pump. We separate parallel and orthogonal components of the probe electric field with a polarizing beamsplitter (BS) after interaction with the sample. Our method ensures temporal synchronization of the different signals, as well as maximum spatial overlap for both measurements. This is crucial, since we target the quantitative comparison of the signals for both polarizations.
\begin{figure}
\includegraphics[width=\columnwidth]{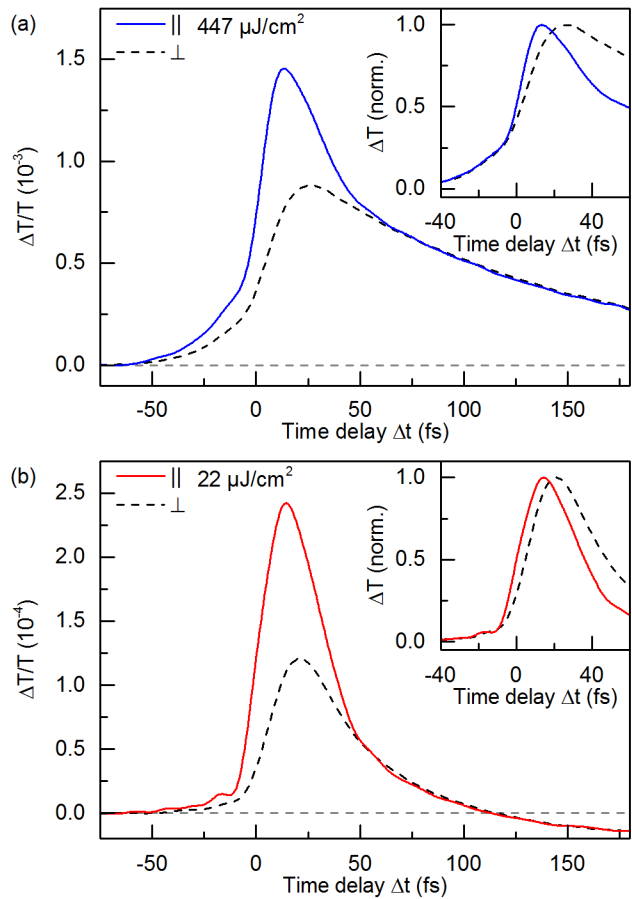}
\caption{\label{Fig3} Relative transmission change $\Delta T/T$ after optical excitation, probed with parallel ($\parallel$) and orthogonal ($\perp$) polarization. Representative measurements for (a) high and (b) low fluences. Insets: signals normalized to unity.}
\end{figure}
\begin{figure}
\includegraphics[width=\columnwidth]{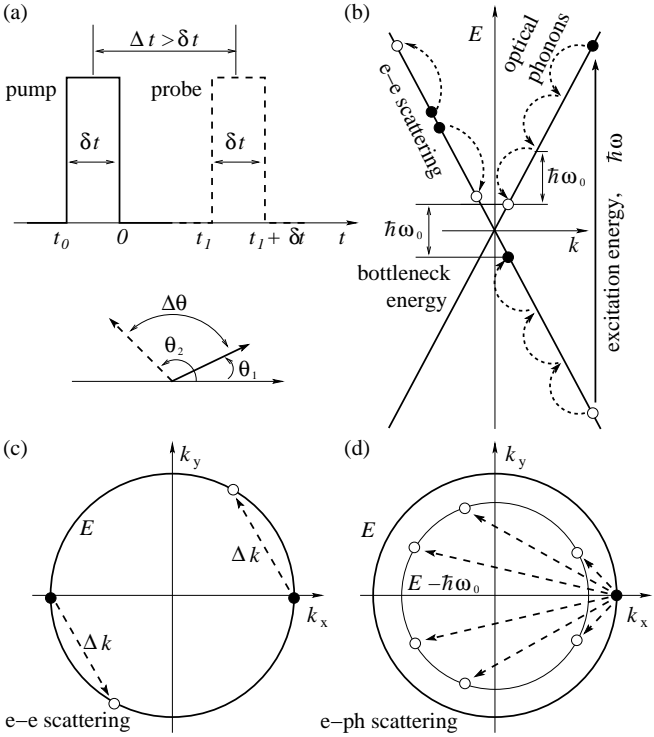}
\caption{\label{Fig2} Fundamental assumptions of our model. (a) Pump and probe pulses are approximated by rectangular profiles. They share the same spectrum and duration but different polarizations, described by the angles $\theta_1$ and $\theta_2$ with $\Delta\theta=\theta_2-\theta_1$. The time delay between the two pulses is $\Delta t$. The model is valid in the case of non-overlapping pulses. (b) The photoexcited carriers relax towards the neutrality point by emitting optical phonons with energy $\hbar\omega_0$. The electron cooling becomes inefficient at the bottleneck energy near $\hbar\omega_0/2$. E-e scattering does not change the total energy of the electronic subsystem. (c) Two-electron scattering across the Dirac cone shown schematically in cross-section at energy $E$. (d) Optical phonons scatter the electrons into any direction. A lower-lying cross section of the Dirac cone at energy $E-\hbar\omega_0$ is shown by a thinner circle.} 
\end{figure}

The differential transmissions for parallel and perpendicular probe components are in Fig.\ref{Fig3}. Excited carriers result in photobleaching of the direct transitions, with an onset time as fast as the duration of the pump pulse. When probing the orthogonal polarization we only observe a signal that, at early times, is significantly weaker, as compared to the configuration with parallel polarizations. We assign this difference to the anisotropy of the electron distribution in the Dirac cone, resulting from the pseudospin selective excitation probability. Already at short time delays$<$50fs the two signals start to converge. The two probe directions become indistinguishable after$\sim$60fs, indicating that the carriers' distribution becomes isotropic in the $k_x-k_y$ plane, while the thermalization along the cone continues.

Importantly, the time required for the distributions to become isotropic does not depend on the excitation fluence. Pure e-e scattering would show a stronger dependence on the starting electron density, with increased probability of e-e collisions\cite{2013winzer}. Indeed, e-e scattering across the Dirac cone 
occurs with a reduced phase space because each electron requires a companion with opposite momentum, as shown in Fig.\ref{Fig2}c. This process is enhanced at higher excitation densities when more scatterers are available. In contrast, the probability for phonon emission does not depend on carrier concentration, when the excitation density is far from saturation.
The optical phonons can scatter the electrons into any direction reducing their energy by $\hbar\omega_0$ and providing conservation for any momentum, see Fig.\ref{Fig2}d,
because the optical branches show limited dispersion in the $k$-space\cite{PRL2004piscanec}.
This, combined with the characteristic timescale of a few tens of fs, allows us to identify e-ph scattering as the main isotropization process.
\begin{figure}
\includegraphics[width=\columnwidth]{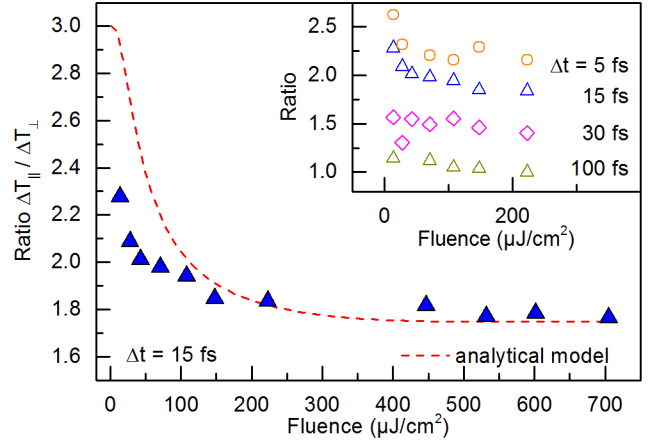}
\caption{\label{Fig4} Ratio of relative transmission changes for parallel and perpendicular probe polarizations for a pump-probe delay of 15fs, when the pulses are negligibly overlapping in time ($\Delta t>\delta t$), but the occupation anisotropy is still substantial ($\Delta t < \tau$), see Fig.\ref{Fig2} for notations.
Experiments (blue triangles) are in good agreement with Eq.(\ref{main}) (red line). The inset is the experimental ratio at time delays between 5 and 100fs.}
\end{figure}

To fully track the dynamics we perform a set of fluence-dependent measurements in which the excitation density is varied between 14 and 705$\mathrm{\mu J/cm^2}$,
see Fig.\ref{Fig4}. We evaluate the ratio between signals for parallel and perpendicular polarization at different pump-probe delays $\Delta t\leq 100$fs. On this time scale, we observe photobleaching, due to strongly nonequilibrium and hot (thermalized) carriers, with just the former being anisotropic. The ratio has a strong dependence on the carrier density for fluences$<$100$\mathrm{\mu J/cm^2}$, i.e. below $10^{13}\,\mathrm{cm}^{-2}$ carriers. Increasing excitation fluence, the isotropic contribution becomes dominant, and we observe a smaller difference between the signals arising for opposite polarizations.

The experiments also show that the maximum differential signal for perpendicular probe photons does not coincide with that for parallel. This depends on fluence and is assigned to the combination of scattering processes along the cone driven by e-e collisions, and momentum isotropization driven by phonons\cite{APL2012malic}. The photoexcited carriers are first redistributed across the Dirac cone due to the optical phonon emission, Fig.\ref{Fig2}d, then along the cone in the new momentum direction, mainly via e-e scattering, Fig.\ref{Fig2}b. The experimental fingerprints of these phenomena are seen in Fig.\ref{Fig3}(insets), where the $\Delta T/T_0$ maxima for different probe polarization do not coincide, having a$\sim$10fs delay. This helps once more to unveil the different role of e-e and e-ph scattering in the overall relaxation dynamics.

Our experiments also show that, at pump fluences$<$50$\mathrm{\mu J/cm^2}$, SLG has increased absorption, i.e. the transient signal becomes negative after$\sim100$fs, as for 
Fig.\ref{Fig3}(b). Refs.\cite{PRL2014kadi,PRB2013sun} assigned this to phonon-assisted intraband absorption of the probe photons. Its contribution is stronger for lower excitation densities, where the weight of intraband transitions of a few hot electrons can be significant\cite{PRL2014kadi,PRB2013sun}. At higher pump intensities, this is hidden under the dominating interband absorption involving a dense hot electron distribution.

{\em Model ---}Charge carriers in graphene near the \textbf{K} point can be described by the massless Dirac Hamiltonian\cite{Nature2007geim} $H_0=\hbar v_0 \hat{\sigma} \cdot \bf{k}$, where $\hbar\bf{k}$ is the two-component momentum operator, $\hat{\sigma}$ is the pseudospin operator derived from the Pauli matrices, and $v_0 = 1.05\cdot 10^6$ms$^{-1}$\cite{NatPhys2011elias}.
The pseudospin orientation in the eigenstates of $H_0$ depends on the direction of $\bf{k}$. It is shown in Fig.\ref{Fig1} as green arrows. The photon-carrier interaction is described by the Hamiltonian $H_\mathrm{int}= \frac{e v_0}{c}\hat{\sigma}\cdot \mathbf{A}$, where $\mathbf{A}=\mathbf{A}_0\cos(\omega t- q z)$ is the vector potential created by the linearly polarized electromagnetic wave ${\mathbf E}=\mathbf{E}_0\sin(\omega t- q z)$ with $\omega=2\pi c/\lambda$ the radiation frequency, and $\mathbf{E}_0=\frac{\omega \mathbf{A}_0}{c}$. We assume normal incidence $\mathbf{q}\perp \mathbf{k}$ without momentum transfer from photons to electrons. Note that the interaction Hamiltonian $H_\mathrm{int}$ inherits the pseudospin dependence from $H_0$ and constitutes the interband transition selection rule of Fig.\ref{Fig1}.

We derive the photocarrier generation rate from the Liouville-von-Neumann equation\cite{VaskoBook} for the density matrix written in the eigenstate basis of $H_0$. The kinetic equation is solved within the duration of the pump pulse $t_0\leq t \leq 0$, and subsequently in the absence of excitation photons for $t>0$ employing
the relaxation time approximation and the solution at $t=0$ as an initial condition, see Fig.\ref{Fig2}a and Appendix. In order to quantify the contribution of the isotropic part of the photocarrier distribution, we assume optical phonon emission as the main mechanism responsible for decay of anisotropy within the first tens fs\cite{APL2012malic}. The electrons are relaxing to the bottleneck energy, see Fig.\ref{Fig2}b, at which further cooling is strongly suppressed. Theory\cite{PRB2007rana} and THz measurements\cite{Nanolett2008george} suggest that the e-h recombination is a much slower process (ps time scale\cite{PRB2007rana,Nanolett2008george}), therefore, the photocarrier concentration is assumed constant. These approximations allow us to write the energy balance equation with only one free parameter, the hot electron temperature, and solve it analytically.

The absorption coefficient $A$ is defined as the ratio of absorbed to incident intensity. In the excited state we sum $A_\tau$  and $A_H$, which describe the absorption due to strongly nonequilibrium carriers and hot electrons, respectively, and $A_0$ denotes the equilibrium value. These three quantities are given in Appendix. With negligible reflection\cite{Science2008nair}, the optical transmissions read $T=1-A$ and $T_0=1-A_0$, resulting in a differential transmission $\frac{T-T_0}{T_0}=-\frac{A-A_0}{A_0} \frac{A_0}{1- A_0}$. The differential absorption is:
\begin{eqnarray}
\nonumber \frac{A-A_0}{A_0} & = & - \left(1+2\cos^2\Delta\theta\right)
 \frac{2 A_0 v_0^2   \Phi_1 }{\hbar\omega^2 \delta\omega^2 \delta t } \mathrm{e}^{-\frac{\Delta t}{\tau}}\\
 &&
 +\tanh\left(\frac{\hbar\omega}{4 T_H} \right) -1,
 \label{main}
\end{eqnarray}
where $A_0=\pi\alpha$ with $\alpha=\frac{e^2}{\hbar c}$ the fine structure constant, $\omega$ the radiation frequency, $\delta\omega$ the spectral width of the pulse,
$\Phi_1$ the pump fluence, $\Delta t$ the probe time delay, $\delta t$ the pulse duration, and $\tau\sim 30\,\mathrm{fs}$ the ``orientational'' relaxation time deduced
from Fig.\ref{Fig3}, where the two curves meet at a time delay $\sim$50fs. The rate of e-e scattering would increase with $\Phi_1$ and diminish $\tau$. The hot carrier temperature, $T_H$, can be estimated as (see Appendix):
\begin{equation}
 \label{TH}
 T_H=\left(\frac{\pi^2 \alpha \hbar^2 v_0^2 \omega_0 \Phi_1}{6\zeta(3)\omega}\right)^{\frac{1}{3}},
\end{equation}
with $\zeta$ the Riemann zeta-function\cite{PrudnikovBook}. The limit of $\Phi_1\to\infty$ is excluded in our model as we assume a system with linear response in $\Phi_1$. $T_H$ cannot be defined until a significant fraction of the strongly nonequilibrium carriers relaxes to the hot Fermi-Dirac distribution\cite{PRB2013tomadin}, i.e. Eqs. \ref{main},\ref{TH} are not valid at $\Delta t=0$. Eq.\ref{main} does not formally vanish for $\Delta t \gg \tau$, because we do not take into account the energy dissipation at such a long time scale. Within the temporal constraints of the model, the description of processes occurring at $\Delta t < \delta t$ is limited. In particular, the slight delay in the maximum of the pump-probe signal, as shown in the inset of Fig.\ref{Fig2}, cannot be reproduced by Eq.(\ref{main}). This does not hinder the description of the isotropization dynamics, experimentally seen over 50fs.

Eq.(\ref{main}) predicts a relative differential transmission$\sim10^{-4}$-$10^{-3}$ at pump fluences between $10$ and $100\,\mathrm{\mu J/cm^2}$, in agreement with the measurements in Fig.\ref{Fig3}. We also evaluate the ratio between differential transmissions for parallel and perpendicular polarizations, and plot this as a function of pump fluence in Fig.\ref{Fig4}. The ratio decreases at higher fluences because the relative contribution of the polarization-dependent term in Eq.(\ref{main}) is strongly suppressed at higher $T_H$, making the overall expression less sensitive to $\Delta \theta$. Physically, the higher excitation density delivers more heat to the isotropic Fermi sea, whose contribution suppressed the strongly nonequilibrium anisotropic component.

{\em Conclusions---} We employed fluence-dependent and polarization-resolved optical pump-probe spectroscopy to resolve and explore the dominating relaxation mechanism for photocarriers in graphene at ultra-short times scales. We found that optical phonon emission, rather than e-e scattering, is responsible for momentum isotropization and pseudospin relaxation in the non-equilibrium photocarrier occupation, while the initial photocarrier redistribution along the Dirac cone in a timescale of tens fs is due to e-e scattering. We provided an analytical framework for the qualitative understanding of the carrier dynamics. Ref.\cite{Nanolett2014echtermeyer} suggested that the light generated anisotropic distribution of carriers in momentum space can be observed in electrical measurements despite their relaxation on ultra-fast time scales, as recently reported in Ref.\cite{Natnano2015tielrooij}. Our model explains why it is possible: the continuous wave laser in Ref.\cite{Nanolett2014echtermeyer} results in low 
photocarrier densities and strong anisotropy, allowing the pseudospin-polarized photocarriers to be detected in graphene pn-junctions\cite{Nanolett2014echtermeyer}. The development of an analytical framework for the description of anisotropy dynamics in graphene paves the way for the qualitative design of novel photodetectors.
\begin{acknowledgments}
We acknowledge  the Emmy Noether Program of the Deutsche Forschungsgemeinschaft (DFG), Zukunftskolleg and EC through the Marie Curie CIG project ``UltraQuEsT'' no. 334463, the DFG through SFB 767, EU Graphene Flagship (Contract No. CNECT-ICT-604391), ERC Synergy Hetero2D, a Royal Society Wolfson Research Merit Award, EPSRC grants EP/K01711X/1, EP/K017144/1, EP/L016087/1.
\end{acknowledgments}

\section{Appendix}
\subsection{Sample preparation}
A 35$\mu$m Cu foil is first annealed at$\sim$1000$^o$C under a 20sccm flow of H$_2$ for 30minutes, followed by 5sccm of CH$_4$. The H$_2$ and CH$_4$ flows are kept constant for 30 minutes, after which the chamber is left to cool down for$\sim$3 hours. SLG is then transferred onto 170$\mu$m thick glass substrates by a wet etching method\cite{Materials2014bonaccorso,Bonaccorso2010review}. A sacrificial layer of polymethyl-methacrylate (PMMA) is spin coated on one side of the Cu foil. The Cu/SLG/PMMA stack is then left to float on the surface of a solution of ammonium persulfate (APS) in water. APS slowly etches Cu, leaving the PMMA+SLG membrane floating. The membrane is picked up using the target glass substrate and, after drying, the PMMA is removed with acetone.
\subsection{Ultrafast pump-probe setup}
\begin{figure}
\includegraphics[width=\columnwidth]{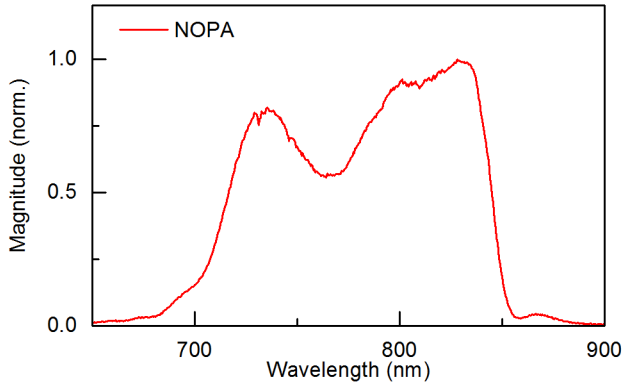}
\caption{\label{Spectrum} Noncollinear optical parametric amplifier output spectrum used in our pump-probe setup.}
\end{figure}

Transient absorption spectroscopy is measured with a Yb:KGW regenerative amplifier system operating at a 50kHz repetition rate. The laser drives a home-built noncollinear optical parametric amplifier (NOPA) which delivers an output spectrum spanning over 0.35eV at a central photon energy of 1.62eV ($\lambda$=765nm, see Fig.\ref{Spectrum})\cite{JoP2010brida}. We compress the pulses to a temporal duration of 15fs by means of chirped dielectric mirrors\cite{JoP2010brida}. The probe pulses are detected with two photodiodes (for parallel and perpendicular polarizations) followed by a lock-in amplifier.
\subsection{Photocarrier generation rate}
We derive the photocarrier generation rate from the Liouville-von Neumann equation\cite{VaskoBook} for the density matrix $\rho$. It is convenient to work in the interaction picture\cite{Cohen-TannoudjiBook}, where the density matrix and light-carrier interaction are given by $\rho^I(t)=\exp(\frac{i}{\hbar}H_0 t)\rho^S(t)\exp(-\frac{i}{\hbar}H_0 t)$, $H_\mathrm{int}^I(t)=\exp(\frac{i}{\hbar}H_0 t) H_\mathrm{int}^S(t)\exp(-\frac{i}{\hbar}H_0 t)$ with $H_0$ the unperturbed Hamiltonian. The superscripts $S$ and $I$ stand for the Schr\"odinger and interaction picture, respectively. The Liouville-von Neumann equation can then be written as\cite{VaskoBook}:
\begin{equation}
\label{L-vN}
\frac{\partial \rho^I}{\partial t}=-\frac{i}{\hbar}[H_\mathrm{int}^I(t),\rho^I(t)].
\end{equation}
Its solution can be obtained using an iteration procedure. First, Eq.(\ref{L-vN}) is integrated to give:
\begin{equation}
\label{L-vNsolution1}
\rho^I(t)-\rho^I(t_0)=-\frac{i}{\hbar}\int\limits_{t_0}^t dt' [H_\mathrm{int}^I(t'),\rho^I(t')].
\end{equation}
Here, $t_0$ has the meaning of initial time, when the interaction is on. Second, Eq.(\ref{L-vNsolution1}) is inserted back to the right-hand side of Eq.(\ref{L-vN}) to give:
\begin{eqnarray}
\nonumber &&
\frac{\partial \rho^I}{\partial t}=-\frac{i}{\hbar}[H_\mathrm{int}^I(t),\rho^I(t_0)]\\
&& - \frac{1}{\hbar^2}\int\limits_{t_0}^t dt' [H_\mathrm{int}^I(t),[H_\mathrm{int}^I(t'),\rho^I(t')]].
\label{L-vNsolution2}
\end{eqnarray}
It is possible to look for higher-order terms in $H_\mathrm{int}^I$. Here, we restrict ourselves to the expansion up to the second order in $H_\mathrm{int}^I$, corresponding to the linear response in terms of the radiation fluence. Now we transform Eq.(\ref{L-vNsolution2}) back to the Schrodinger picture. The left-hand side of Eq.(\ref{L-vNsolution2}) then reads:
\begin{equation}
\label{lhs}
\frac{\partial \rho^I}{\partial t}\to \frac{\partial \rho^S}{\partial t}+\frac{i}{\hbar}[H_0,\rho^S(t)].
\end{equation}
The right-hand side of Eq.(\ref{L-vNsolution2}) consists of two terms:
\begin{equation}
\label{rhs1}
[H_\mathrm{int}^I(t),\rho^I(t_0)]\to [H_\mathrm{int}^S(t),\mathrm{e}^{-\frac{i}{\hbar}H_0\Delta t}\rho^S(t_0)\mathrm{e}^{\frac{i}{\hbar}H_0\Delta t}]
\end{equation}
and
\begin{eqnarray}
\label{rhs2} &&
\int\limits_{t_0}^t dt' [H_\mathrm{int}^I(t),[H_\mathrm{int}^I(t'),\rho^I(t')]] \to \\
&&\int\limits_{t_0-t}^0 dt'' [H_\mathrm{int}^S(t),\mathrm{e}^{\frac{i}{\hbar}H_0 t''}[H_\mathrm{int}^S(t+t''),\rho^S(t+t'')]\mathrm{e}^{-\frac{i}{\hbar}H_0 t''}],
\nonumber
\end{eqnarray}
where  $t''=t'-t$.
A linearly polarized electromagnetic wave propagating in the $z$-direction with wave vector $q$ and frequency $\omega$ can be described by the vector potential
$\mathbf{A}=\mathbf{A}_0 \cos\left(\omega t -qz \right)$ which relates to the corresponding electric field as $\mathbf{E}= \mathbf{E}_0 \sin\left(\omega t -qz \right)$,
where $\mathbf{E}_0 = \omega \mathbf{A}_0/c$. In the case of normal incidence, $q$ does not have any influence on the in-plane carrier momenta $k$. The interaction Hamiltonian $H_\mathrm{int}^S(t)$ can be deduced from the tight-binding effective Hamiltonian\cite{Nature2007geim} and written as\cite{EPL2012trushin}:
\begin{equation}
\label{Hint}
H_\mathrm{int}^S(t)=h_\mathrm{int} \left(\mathrm{e}^{i\omega t} + \mathrm{e}^{-i\omega t}\right),
\end{equation}
where
\begin{equation}
h_\mathrm{int} =\frac{ev_0 E_0}{2\omega}\left(
\begin{array}{ll}
0 & \mathrm{e}^{-i\theta}\\
\mathrm{e}^{i\theta} & 0
\end{array}\right),
\end{equation}
with $\tan \theta = A_{0y}/A_{0x}$ the polarization angle. Substituting Eq.(\ref{Hint}) into Eqs.(\ref{rhs1},\ref{rhs2}), and neglecting the fast oscillating terms of the form $\mathrm{e}^{\pm i\omega t}$, we obtain the following equation for the density matrix:
\begin{equation}
\label{gph0}
\frac{\partial \rho^S}{\partial t}+\frac{i}{\hbar}[H_0,\rho^S(t)] = g(t),
\end{equation}
where
\begin{eqnarray}
\label{gph} \nonumber &&
g(t) = -\frac{1}{\hbar^2}\int\limits_{t_0-t}^0 dt'' \left\{\left(\mathrm{e}^{i\omega t''} + \mathrm{e}^{-i\omega t''}\right)  \right.\\
&&
\left. \times [h_\mathrm{int},\mathrm{e}^{\frac{i}{\hbar}H_0 t''}[h_\mathrm{int},\rho^S(t+t'')]\mathrm{e}^{-\frac{i}{\hbar}H_0 t''}] \right\},
\end{eqnarray}
is the photogeneration rate. Eqs.(\ref{gph0},\ref{gph}) are the operator equations. To get the corresponding expression for the distribution function we rewrite Eqs.(\ref{gph0},\ref{gph}) in the helicity basis, i.e. the eigenfunction basis of the unperturbed Hamiltonian $H_0$\cite{Nature2007geim}. The resulting distribution function is a $2\times 2$ matrix. We retain only its diagonal elements, relevant for interband optical transitions\cite{VaskoBook}. Thus, instead of Eq.(\ref{gph0}) we get:
\begin{equation}
\label{gph1}
\frac{\partial f_\pm(t)}{\partial t}=g_\pm\left[f_\pm(t)\right],
\end{equation}
where $g_\pm\left[f_\pm(t)\right]$ is given by:
\begin{eqnarray}
\label{gph2} \nonumber &&
g_\pm\left[f_\pm(t)\right]=-\frac{1}{\hbar^2}\int\limits_{t_0 -t}^0 dt'' \left\{ \left[\mathrm{e}^{it''(\omega - \Omega)} + \mathrm{e}^{-i t''(\omega - \Omega)}\right] \right.\\ &&
\left. \times  \left[f_\pm(t+t'')-f_\mp(t+t'')\right] h_{12}h_{21} \right\}.
\end{eqnarray}
Here, $f_\pm(t)$ is the electron distribution function in either valence (subscript ``$-$'') or conduction (subscript ``$+$'') band, $\hbar \Omega=2\hbar v_0 k$ is the interband transition energy, and $h_{ij}$ are the matrix elements of $h_\mathrm{int}$ given by:
\begin{equation}
h_{ij} =\frac{ev_0 E_0}{2\omega}\left(
\begin{array}{ll}
\cos(\phi-\theta) & -i\sin(\phi-\theta)\\
i\sin(\phi-\theta) & \cos(\phi-\theta)
\end{array}\right),
\end{equation}
where $\tan\phi=k_y/k_x$. The product $h_{12}h_{21}$ can be thus written as:
\begin{equation}
\label{probab}
h_{12}h_{21} = \left(\frac{ev_0 E_0}{2\omega}\right)^2\sin^2(\phi-\theta).
\end{equation}
Eq.~(\ref{probab}) reflects the polarization dependency in the photocarrier generation rate given by Eq.~(\ref{gph2}).
\subsection{Description of the pump pulse}
The pump pulse excites the carriers over the Fermi sea described by the Fermi-Dirac distribution $f_\pm^{(0)}$. The generation rate then derives from Eq.~(\ref{gph2}):
\begin{eqnarray}
\label{gphpump} \nonumber &&
g_\pm^{(1)}= -\frac{2 h_{12}^{(1)}h_{21}^{(1)}}{\hbar^2} \frac{\sin\left[(t-t_0)(\omega_1 - \Omega)\right]}{\omega_1 - \Omega} \\
&&
\times  \left(f_\pm^{(0)}-f_\mp^{(0)}\right), \quad t>t_0
\end{eqnarray}
where the pulse is switched on at $t_0<0$, see Fig.\ref{Fig2}.
The evolution of the non-equilibrium distribution function within the pump pulse can be written as:
\begin{equation}
\label{eq1}
\frac{\partial f_\pm^{(1)}}{\partial t}=g_\pm^{(1)} - \frac{f_\pm^{(1)}}{\tau},\quad t_0<t<0
\end{equation}
where $\tau$ is the relaxation time, and $f_\pm^{(1)}$ reads:
\begin{eqnarray}
\label{sol1} &&
f_\pm^{(1)}=\frac{\tau(\omega_1 - \Omega)}{1+\tau^2(\omega_1 - \Omega)^2} \frac{2 h_{12}^{(1)}h_{21}^{(1)}}{\hbar^2 (\omega_1 - \Omega)^2}  \left(f_\pm^{(0)}-f_\mp^{(0)}\right)\\
\nonumber &&
\times  \left\{\tau(\omega_1 - \Omega) \cos[(\omega_1 - \Omega)(t-t_0)]-\sin[(\omega_1 - \Omega)(t-t_0)]\right\},\\
\nonumber &&
\qquad t_0 < t < 0.
\end{eqnarray}
One can prove by direct substitution that $f_\pm^{(1)}$, given by Eq.(\ref{sol1}), satisfies Eq.~(\ref{eq1}).

Once the pump pulse is switched off at $t=0$, Eq.~(\ref{eq1}) becomes:
\begin{equation}
\label{eq2}
\frac{\partial f_\pm^{(1)}}{\partial t}= - \frac{f_\pm^{(1)}}{\tau},\quad t>0
\end{equation}
Its solution is $f_\pm^{(1)}(t)=f_\pm^{(1)}(0)\mathrm{e}^{-t/\tau}$, where $f_\pm^{(1)}(0)$ is given by (\ref{sol1}) at $t=0$. The non-equilibrium distribution function at $t>0$ thus reads:
\begin{eqnarray}
\label{sol2} &&
f_\pm^{(1)}=\frac{\tau(\omega_1 - \Omega)}{1+\tau^2(\omega_1 - \Omega)^2} \frac{2 h_{12}^{(1)}h_{21}^{(1)}}{\hbar^2 (\omega_1 - \Omega)^2}  \left(f_\pm^{(0)}-f_\mp^{(0)}\right)\\
\nonumber &&
\times  \left\{\tau(\omega_1 - \Omega) \cos[(\omega_1 - \Omega)\delta t]-\sin[(\omega_1 - \Omega)\delta t]\right\}\mathrm{e}^{-t/\tau},\\
\nonumber &&
\qquad t>0,
\end{eqnarray}
where $\delta t$ plays the role of the pulse duration. The energy relaxation time is assumed to be much longer than $\tau$. Therefore, the total distribution function at $t>0$ represents the sum of the hot Fermi-Dirac distribution $f_{\pm}^{(H)}$ and the non-equilibrium addition, Eq.~(\ref{sol2}).
	
The prefactor $h_{12}^{(1)}h_{21}^{(1)}$ can be rewritten in terms of pump fluence $\Phi_1=(cE_1^2\delta t)/(8\pi)$ (with $E_1$ the electromagnetic wave amplitude), so that Eq.(\ref{probab}) becomes:
\begin{equation}
h_{12}^{(1)}h_{21}^{(1)}=\frac{2\pi \alpha \hbar v_0^2 \Phi_1}{\omega_1^2 \delta t}\sin^2(\phi-\theta_1),
\end{equation}
where $\alpha=e^2/(\hbar c)$ is the fine structure constant, and $\tan\theta_1 = E_{1y}/E_{1x}$. Since $\delta t$ is much longer than the typical time scale $\omega_1^{-1}$ determined by the optical frequency, we consider the limit of $\delta t \to \infty$. $\tau$ must be set to infinity at the time scale $\omega_1^{-1}$, because $\tau > \delta t$ by definition. Using the formula\cite{Cohen-TannoudjiBook}:
\begin{equation}
\label{delta}
\lim_{\delta t \to \infty}\frac{4\sin^2[(\omega_1 - \Omega)\delta t/2]}{(\omega_1 - \Omega)^2\delta t}=2\pi \delta(\omega_1 - \Omega)
\end{equation}
we get the approximated expression for Eq.(\ref{sol2}):
\begin{eqnarray}
\label{sol3} &&
f_\pm^{(1)}=
\frac{4\pi^2 \alpha v_0^2 \Phi_1}{\hbar \omega_1^2 }\sin^2(\phi-\theta_1) \\
\nonumber && \times  \left(f_\mp^{(0)}-f_\pm^{(0)}\right) \delta(\omega_1 - \Omega) \mathrm{e}^{-t/\tau}, \quad t>0.
\end{eqnarray}
\subsection{Description of the probe pulse}
We now consider the optical absorption $A=A_H + A_\tau$ of the probe pulse governed by the hot electrons (index $H$) and strongly non-equilibrium short-living photocarriers (index $\tau$) created by the pump pulse. We first calculate $A_0$ in the absence of the pump pulse. The total number of the optical interband transitions $G_0$ within the pulse duration $\delta t$ can be evaluated by integrating the generation rate Eq.(\ref{gph2}) over the time $t$:
\begin{eqnarray}
\label{G0-1} \nonumber &&
G_0=
\frac{2}{\hbar^2}\int\limits_{t_1}^{t_1+\delta t} dt\int\limits_{t_1 -t}^0 dt'' \cos\left[t''(\omega_2 - \Omega)\right] \\
&&
\times   \left(f^{(0)}_- -f^{(0)}_+ \right) h^{(2)}_{12}h^{(2)}_{21} ,
\end{eqnarray}
where $f^{(0)}_\pm$ is the equilibrium carrier distribution function at lattice temperature, and:
\begin{equation}
h_{12}^{(2)}h_{21}^{(2)}=\frac{2\pi \alpha \hbar v_0^2 \Phi_2}{\omega_2^2 \delta t}\sin^2(\phi-\theta_2),
\end{equation}
with $\Phi_2$,  $\omega_2$, $\theta_2$ the probe pulse fluence, frequency, and polarization angle.
Taking the integrals in Eq.(\ref{G0-1}):
\begin{eqnarray}
\label{G0-2}
&&
G_0= \frac{2\pi \alpha v_0^2 \Phi_2 }{\hbar\omega_2^2 }\sin^2(\phi-\theta_2)  \\
\nonumber &&
\times   \frac{4\sin^2\left[(\omega_2 -\Omega)\delta t/2  \right]}{(\omega_2 -\Omega) \delta t } \left(f^{(0)}_- -f^{(0)}_+ \right).
\end{eqnarray}
We again exploit the fact that $\delta t >> \omega_2^{-1}$ and utilize Eq.~(\ref{delta}) for the transformation of the second line of Eq.~(\ref{G0-2}) into the $\delta$-distribution. The absorbed fluence for a given valley/spin channel can be then calculated as:
\begin{eqnarray}
\label{Phi0}
\Phi_0 &=& \int\limits_{0}^{2\pi}\frac{d\phi}{4\pi^2} \int\limits_{0}^{\infty}\frac{d\Omega\Omega}{4 v_0^2} G_0(\Omega, \phi)=\frac{\pi\alpha}{4} \Phi_2
\end{eqnarray}
Here, we assume $f_-^{(0)}=1$ and $f_+^{(0)}=0$ at $\Omega=\omega_2$. $A_0$ should include the spin/valley degeneracy:
\begin{equation}
\label{P0}
A_0=\pi\alpha.
\end{equation}
This result agrees with previous measurements\cite{Science2008nair}.
	
We now consider the probe pulse absorption due to the hot carriers created by the pump pulse. The hot carriers are described by the hot Fermi-Dirac distribution $f_{\pm}^{(H)}$, which should now substitute $f_{\pm}^{(0)}$ in Eqs.(\ref{G0-1},\ref{G0-2}). Since the chemical potential is much smaller than the excitation energy we set former to zero. The occupation difference can then be written as:
\begin{equation}
f^{(0)}_- -f^{(0)}_+ = \tanh\left(\frac{\hbar\Omega}{4 T_H} \right),
\end{equation}
and the absorbed fluence in the presence of hot carriers for a given valley/spin channel as:
\begin{equation}
\Phi_H=\frac{\pi\alpha}{4}  \tanh\left(\frac{\hbar\omega_2}{4 T_H} \right)\Phi_2,
\end{equation}
where $T_H$ is the hot carrier temperature estimated below.
The corresponding optical absorption in the one-color pump-probe setup $\omega_1=\omega_2=\omega$ is then given by:
\begin{equation}
\label{PH}
A_H=\pi\alpha \tanh\left(\frac{\hbar\omega}{4 T_H} \right).
\end{equation}
Finally, we calculate the optical absorption due to the strongly non-equilibrium time-dependent carrier distribution after the pump pulse. The interband transition rate can be evaluated from Eq.(\ref{gph2}):
\begin{eqnarray}
\label{gph-1} \nonumber &&
g _\tau(t)=
\frac{2}{\hbar^2}\int\limits_{t_1 -t}^0 dt''  \cos\left[t''(\omega_2 - \Omega)\right] \\
&&
\times   \left[f^{(1)}_-(t+t'')-f^{(1)}_+(t+t'') \right]  h^{(2)}_{12}h^{(2)}_{21} .
\end{eqnarray}
Here, the out-of-equilibrium distribution $f^{(1)}_\pm$ is given by either Eq.(\ref{sol2}) or (\ref{sol3}), depending on the approximation used. In what follows we employ the latter because we have already utilized a somewhat similar approximation to derive $A_0$ and $A_H$. Note that $f_-^{(1)}=-f_+^{(1)}$ because of the SLG e-h symmetry. The number of optical interband transitions within the probe pulse then reads:
\begin{eqnarray}
\nonumber \label{Gtau} && G_\tau =-\frac{4\tau f^{(1)}_+(0)}{\hbar^2 (\omega_2 -\Omega)}  \frac{h^{(2)}_{12}h^{(2)}_{21}}{1+(\omega_2 -\Omega)^2 \tau^2}
\left( (\omega_2 -\Omega)\tau \mathrm{e}^{-\Delta t/\tau} \right.\\
\nonumber && \left.   + \mathrm{e}^{-t_1/\tau}\left\{
\sin\left[\delta t (\omega_2 -\Omega)\right]-(\omega_2 -\Omega)\tau \cos\left[\delta t (\omega_2 -\Omega)\right]\right\}\right),\\
\end{eqnarray}
and the fluence absorbed within this process can be obtained by integrating $G_\tau$ over the whole $k$-space, and by subsequent averaging over the pulse duration. The latter makes the fast oscillating terms in the second line of (\ref{Gtau}) vanish. We thus get:
\begin{equation}
\Phi_\tau= -\frac{\pi^2\alpha^2 v_0^2 \Phi_1 \Phi_2}{2\hbar \omega_1 \omega_2 \delta\omega^2 \delta t}\left[2+\cos\left(2\Delta\theta\right)\right]
\mathrm{e}^{-\Delta t/\tau},
\end{equation}
where $\Delta\theta =\theta_2-\theta_1$, and $\delta\omega=\omega_2-\omega_1$. In the one-color pump-probe setup $\omega_1=\omega_2=\omega$, and $\delta\omega$ plays the role of energy uncertainty (estimated as $\delta t^{-1}$). The resulting probe pulse absorption due to the strongly out-of-equilibrium carriers created by the pump pulse is:
\begin{equation}
\label{Ptau}
A_\tau =-\frac{2\pi^2\alpha^2 v_0^2 \Phi_1}{\hbar \omega^2 \delta\omega^2 \delta t}\left(1+2\cos^2\Delta\theta\right)
\mathrm{e}^{-\Delta t/\tau}.
\end{equation}
\section{Hot temperature calculation}
To estimate the hot electron temperature as a function of the pump pulse we assume that (i) the e-h recombination process is much slower than $\tau$ and, therefore,
the photocarrier concentration can be considered as a constant on a time scale of a few tens fs; (ii) the nonequilibrium photocarrier occupation relaxes towards the hot Fermi-Dirac distribution mostly due to the optical phonon emission with the frequency $\omega_0$, resulting in a characteristic electron energy $\sim\hbar\omega_0/2$ after thermalization, see Fig.\ref{Fig2}.
The photoelectron concentration $n_\mathrm{ph}$ for a given spin/valley channel can be written as\cite{PRB2007rana}:
\begin{equation}
 \label{n}
 n_\mathrm{ph}=\int\frac{d^2k}{4\pi^2}f_+^{(1)},
\end{equation}
whereas for photoholes we have\cite{PRB2007rana}:
\begin{equation}
 \label{p}
 p_\mathrm{ph}=\int\frac{d^2k}{4\pi^2}\left(1-f_-^{(1)} \right).
\end{equation}
Here, $f_\pm^{(1)}$ is given by Eq.(\ref{sol3}). The energy balance equations for a given spin/valley channel become:
\begin{equation}
 \label{energy-n}
 n_\mathrm{ph}\frac{\hbar\omega_0}{2}+\int\frac{d^2k}{4\pi^2} \hbar v_0 k f_+^{(0)}=\int\frac{d^2k}{4\pi^2} \hbar v_0 k f_+^{(H)},
\end{equation}
\begin{equation}
\label{energy-p}
p_\mathrm{ph}\frac{\hbar\omega_0}{2}+\int\frac{d^2k}{4\pi^2} \hbar v_0 k (1-f_-^{(0)})=\int\frac{d^2k}{4\pi^2} \hbar v_0 k (1-f_-^{(H)}).
\end{equation}
Here $f_\pm^{(0)}$ ($f_\pm^{(H)}$) is the Fermi-Dirac distribution at the lattice (hot) temperature. The integrals in Eqs.(\ref{n}-\ref{energy-p}) can be solved analytically in the case of intrinsic SLG, i.e. at zero doping. However, our samples are not intrinsic with a doping of the order of 100meV.
To proceed, we assume that the chemical potential $\mu>0$ is higher than the lattice (room) temperature $T_L$,
but lower than the hot electron temperature $T_H$. For electrons, we have:
\begin{equation}
 \label{int1}
 \int\frac{d^2k}{4\pi^2} \hbar v_0 k f_+^{(H)} \approx \frac{3}{4\pi} \frac{\zeta(3) T_H}{\hbar^2 v_0^2},
\end{equation}
\begin{equation}
 \label{int2}
 \int\frac{d^2k}{4\pi^2} \hbar v_0 k f_+^{(0)}  \approx \frac{\mu^3}{6\pi\hbar^2 v_0^2},
\end{equation}
and the electron energy balance reads:
\begin{equation}
 \label{energy2}
 \frac{\pi\alpha\omega_0 \Phi_1}{8\omega_1} +\frac{\mu^3}{6\pi\hbar^2v_0^2}=\frac{3}{4\pi} \frac{\zeta(3) T_{H}^3}{\hbar^2 v_0^2}.
\end{equation}
Here $\zeta$ is the Riemann $\zeta$-function\cite{PrudnikovBook}.
The hot electron temperature is given by:
\begin{equation}
 \label{THdoped}
 T_H=\left(\frac{\pi^2 \alpha \hbar^2 v_0^2 \omega_0 \Phi_1}{6\zeta(3)\omega_1}+ \frac{2\mu^3}{9\zeta(3)}\right)^{\frac{1}{3}}.
\end{equation}
The holes at $\mu<0$ can be considered in a similar way.
Assuming typical values $\Phi_1=10-100\,\mathrm{\mu J/cm^2}$, $\mu$ of the order of $0.1\,\mathrm{eV}$ we find that $T_H$ ranges from 800 to 1800 K.
The $\mu$-dependent term in Eq.(\ref{THdoped}) contributes weakly to $T_H$ and can be neglected. 
The hot electron and hole temperatures are equal within this approximation.
Physically, intrinsic electrons, while being at lattice temperature, do not contribute much to the energy balance, even though their concentration might be high. 
Thus, we arrive at Eq.~(\ref{TH}).

\end{document}